\documentclass[aps,prl,twocolumn,citeautoscript,superscriptaddress]{revtex4-1}

\usepackage{color}
\usepackage{graphicx}
\usepackage{hyperref}

\begin{document}

\title{
Magnetic field inside of a nucleon and the quark-Zeeman effect
}
\author{Marek Matas}
\affiliation{Faculty of Nuclear Sciences and Physical Engineering,
Czech Technical University in Prague, Czech Republic. Correspondence and requests for materials should be addressed to matas.marek1@gmail.com.}


\begin{abstract}
Zeeman effect has been an invaluable tool for the detection and measurement of the effects of a magnetic field on the internal behaviour of atoms and nuclei. Along with M\"{o}ssbauer effect, it has been used to determine the magnitude of the magnetic field present in nuclei of atoms but has not been detected inside nucleons so far. In this work, we present a possible signature of the Zeeman-like effect in nucleons that one can extract from the parton distributions with the use of statistical models and Fermi-Dirac distribution. These effects in turn---if electromagnetically dominated---predict the presence of magnetic fields of an unprecedented magnitude inside protons.
\end{abstract}


\maketitle

\section{\label{sec:Intro}Introduction}
In the theory of standard model, nucleons (protons and neutrons) consist of fermionic fields called quarks and of a number bosonic fields that bind them together (gluons, photons, etc.). These particles forming a nucleon are probed and studied in numerous ways---one of the most successful methods to study the internal structure of a nucleon are the lepton-hadron collisions called deep inelastic scattering. A usual way of describing these particles---called partons---is in terms of the so-called parton distribution functions (PDFs). These distributions tell us how many partons carry a certain momentum at a given scale and, due to their universality, can be used in any factorized cross section formula.

Statistical approach to PDFs was proposed and used to describe the measured helicity-dependent quark distributions in terms of a few phenomenological terms and a Fermi-Dirac distribution~\cite{Bourrely:2013yti, Bourrely:2015kla,Bourrely:2018yck}, where helicity refers to the alignment of the spin of the considered quark with the spin of the hadron that contains it. The Fermi-Dirac distribution that is used to model the behaviour of highly-energetic quarks contains a dimensionless thermodynamic potential-like variable. This potential reaches different values for different helicities of the modelled quark as well as for different flavors.

In this work, we link the different values of the internal quark-potential with the manifestation of the Zeeman effect---energy level splitting induced by the presence of a magnetic field---and then obtain an estimate for the magnitude of the magnetic field that is present at deep inelastic scattering events in nucleons.

\section{\label{sec:statpdf}Statistical parton distributions}
\begin{figure}[h!]
  \centering
   \includegraphics[width=\linewidth]{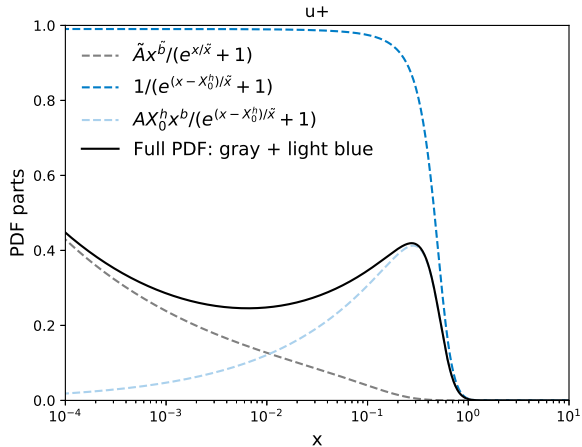}
\caption{\textbf{Constituent terms of the statistical form} \textbf{of a PDF for an u-quark with positive helicity.} In this plot we see the dissection of the statistical u-quark distribution into its constituent terms. We can see that the term that is sensitive to the quark potential energy (dashed light blue line) is dominant in the high-energy region of the proton---at high values of the carried fractional momentum $x$.}
\label{fig:u+}
\end{figure}

Statistical parton distribution approach aims to model the universal quark and gluon parton distribution functions in term of a few phenomenologically-motivated terms and a Fermi-Dirac distribution. In the work of Bourrely and Soffer~\cite{Bourrely:2015kla}, helicity-dependent quark distributions have been modelled with the expression
\begin{equation}\label{eq:stat_pdf}
xq^h(x, Q_0^2) = \frac{AX_{0q}^h x^b_q}{\exp{[(x - X_{0q}^h)/\bar{x}]} + 1} + \frac{\tilde{A} x^{\tilde{b_q}}}{\exp{(x / \bar{x})} + 1},
\end{equation}
where the first term contains helicity-dependent Fermi-Dirac distribution modelling the highly-energetic quark contribution. The second term models the low-$x$ growth of the PDFs, where $x$ is the fractional momentum carried by the parton, (as can be seen from Fig.~\ref{fig:u+} for the $u$-quark distribution with positive helicity). Free variables of these distributions have been fitted at $Q_0^2$ = 1\,GeV$^2$ to values of 
\begin{equation}
    \bar{x} = 0.090 \pm 0.002
\end{equation}
for the universal nucleon thermodynamical temperature and
\begin{eqnarray}\label{eq:parameter}
    X_{0u}^+ & = 0.475 \pm 0.001, \hspace{20pt} X_{0u}^- = 0.307 \pm 0.001, \nonumber \\ 
    X_{0d}^+ & = 0.245 \pm 0.001, \hspace{20pt} X_{0d}^- = 0.309 \pm 0.001,\\ \nonumber
    X_{0s}^+ & = 0.011 \pm 0.001, \hspace{20pt} X_{0s}^- = 0.015 \pm 0.001,
\end{eqnarray}
for the different values of thermodynamical potentials~\cite{Bourrely:2015kla}. Subscript denotes the flavor of the considered quark; superscript is positive when helicity of the quark is aligned in the same direction as the helicity of the nucleon and negative otherwise. The fit used combined data from unpolarized deep inelastic scattering (both neutral and charge current) arriving to a $\chi^2$ of 2288 for 1773 data points and polarized deep inelastic scattering arriving to a $\chi^2$ of 319 for 269 data points. Values of other parameters as well as their interpretation are not relevant for the content of this paper and are accessible in~\cite{Bourrely:2015kla}. 

We notice that the change of the potential with respect to helicity flip is qualitatively in accordance with the Zeeman effect, namely that going from positive to negative helicity decreases the potential of the $u$ quark (positive charge) and increases for $d$ and $s$ quarks (negative charge). We also observe, that the change in potential is ordered in the same manner as quark rest-mass, which is also one of the signatures of the Zeeman effect as discussed in next sections.

\section{\label{sec:formula}Generalizing the Zeeman-effect formula}
Zeeman effect describes the amount of perturbation induced in the potential energy of a particle when exposed to an external static magnetic field. From Fig.~\ref{fig:u+}, we can see that the term that is sensitive to the quark-potential energy lies in the high-energy region of the proton. That is why we can assume linear form of the Zeeman-potential perturbation $\Delta U$ as
\begin{equation}\label{eq:electronzeem}
    \Delta U = -\vec{\mu} \vec{B}
\end{equation}
and do not assume non-linear effects that would be present for the sea quark-antiquark pair distributions that are expected to be in a bound J$^{\mathrm{PC}}$ 0$^{++}$ state (as discussed by Sauder and Desolates for the case of positronium~\cite{Sauder}). $\vec{B}$ is the magnetic field and the magnetic moment $\vec{\mu}$ can be expressed as
\begin{equation}\label{eq:mu}
    \vec{\mu} = -\frac{\mu_B (g_l \vec{L} + g_s \vec{S})}{\hbar},
\end{equation}
where $\mu_B$ is the so-called Bohr magneton, $\vec{L}$ is the orbital angular momentum, $\vec{S}$ is the spin angular momentum and $g_{l,s}$ are the Land\'{e} $g$-factors for angular momentum and spin interactions. Spin-orbital momentum has been shown to have twice the influence on the magnetic moment of a particle than the angular orbital momentum and so $g_l$ = 1 and $g_s\approx$ 2. This is a consequence of linking the Bohr magneton with magnetic dipole moment obtained from the Dirac equation and it holds both for electrons and quarks.

Since we are interested in the change of the potential energy of a quark with respect to the change of its helicity, we are only interested in the projection of the magnetic field onto the $z$-axis (in the direction of the flight of the nucleon) and so we can rewrite Eq.~(\ref{eq:electronzeem}) as

\begin{equation}\label{eq:electronzeem_true}
    \Delta U = -|\vec{\mu}| B_z.
\end{equation}
Furthermore, since we assume identical orbital momentum distributions and solely a helicity flip of the studied quark, the orbital angular momentum is averaged out and we can omit this term in Eq.~(\ref{eq:mu}) as
\begin{equation}\label{eq:zeeman}
    |\vec{\mu}| = -\frac{\mu_B g_s |\vec{S}|}{\hbar} = -\frac{2 \mu_B}{\hbar},
\end{equation}
where we have utilized that we are expressing a difference between positive and negative helicity of a spin-1/2 particle so the total spin change is 1 and also that the electron spin $g$-factor is $\sim$ 2.

The fact, that Eq.~(\ref{eq:electronzeem_true}) is independent of the orbital momentum is also the reason why we are not calling the effect of energy-level splitting the Paschen-Back effect that would otherwise be the dominant force in the presence of strong magnetic fields inside nucleons. Since the orbital momentum dependence is washed away, the two effects become identical and for historical reasons, we will stick to the term Zeeman effect.

Bohr magneton is defined for electrons as
\begin{equation}\label{eq:bohrmagneton}
    \mu_B = \frac{e \hbar}{2m_e},
\end{equation}
where $e$ is the elementary charge, $m_e$ is the electron mass and $\hbar$ is the reduced Planck constant. We will redefine the Bohr magneton for the quark case as
\begin{equation}\label{eq:bohrmagnetonquark}
    \mu_B^f = \frac{e q_f \hbar}{2m_f},
\end{equation}
where $q_f$ is the fractional charge of a quark of flavor $f$ and $m_f$ is its mass. These values have been set as 
\begin{eqnarray}\label{eq:masses}
    m_u & = & 2.16^{+0.49}_{-0.26} \hspace{5pt} \text{MeV}, \hspace{20pt} q_u = 2/3, \nonumber \\ 
    m_d & = & 4.67^{+0.48}_{-0.17} \hspace{5pt} \text{MeV},  \hspace{20pt} q_d = -1/3,\\ \nonumber
    m_s & = & 93^{+11}_{-5} \hspace{5pt} \text{MeV}, \hspace{35pt} q_s = -1/3,
\end{eqnarray}
as taken from~\cite{Tanabashi:2018oca}.

\section{\label{sec:Scale}Bringing mass-scale into the statistical distribution}
The statistical parton distribution approach~\cite{Bourrely:2015kla} uses dimensionless parameters to represent the universal proton temperature and quark potential. This then in turn prevents us to use these potentials directly in the Zeeman effect formula (Eq.~(\ref{eq:electronzeem_true})) because here we need to identify the potential shift with a dimensionful quantity. This cannot be done directly in the framework of this model.

However, there are other models of statistical parton distributions where they modelled the valence contribution to the quark wave functions by the Fermi-Dirac distributions with mass-dependent parameters~\cite{Zhang:2008nr, Sohaily:2017etw}. These works do not include the helicity-dependent potential fits that we have described in the previous sections and that are necessary for the computation of the quark-Zeeman effect. They, however, fit the universal proton temperature in terms of a dimensionful quantity rather than as a fractional dimensionless value.

As an example, in the work of Zhang et.~al.~\cite{Zhang:2008nr}, they arrive at a universal proton temperature of $T_0$ = 47\,MeV and in the work of Sohaily and Vaziri~\cite{Sohaily:2017etw}, it is set as $T_0$ = 51.5\,MeV.

We can use the values of the dimensionful universal proton temperature that was computed in the unpolarized framework to get a rough estimate of the scale with which we can restore dimension to the helicity-dependent potentials as
\begin{equation}\label{eq:restorescale}
    U_{0q}^h = X_{0q}^h E_{res},
\end{equation}
where
\begin{equation}\label{eq:restorescale2}
    E_{res} = \frac{T_0}{\bar{x}}.
\end{equation}
We have set $T_0$ = 50\,MeV and $\bar{x}$ as 0.09 as described above arriving finally at
\begin{eqnarray}\label{eq:parameterscale}
    U_{0u}^+ & = 263.89 \hspace{5pt} \text{MeV}, \hspace{20pt} U_{0u}^- = 170.56 \hspace{5pt} \text{MeV}, \nonumber \\ 
    U_{0d}^+ & = 136.11 \hspace{5pt} \text{MeV}, \hspace{20pt} U_{0d}^- = 171.67 \hspace{5pt} \text{MeV},\\ \nonumber
    U_{0s}^+ & = 6.11 \hspace{5pt} \text{MeV}, \hspace{20pt} U_{0s}^- = 8.33 \hspace{5pt} \text{MeV}.
\end{eqnarray}

\section{\label{sec:magfield}Magnetic field inside of a nucleon}
\begin{figure}[h]
  \centering
   \includegraphics[width=\linewidth]{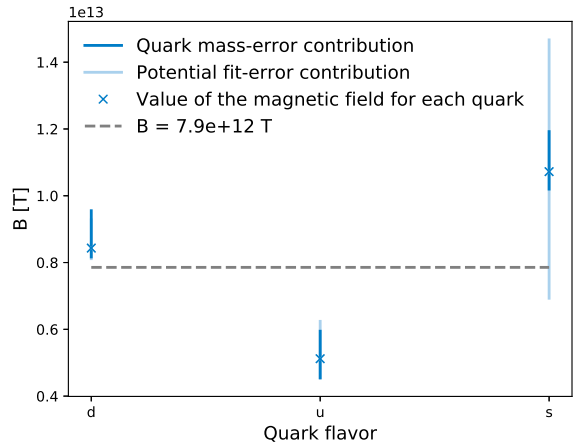}
\caption{\textbf{Magnitude of the magnetic field inside of a nucleon computed for various quark distributions.} In this plot we see the values of the magnetic field obtained from the Zeeman-induced potential energy perturbation for the three light-quark flavors u,d, and s. The fitted value of the potential reaches a value of $B=7.9 \cdot 10^{12}$\,T and is shown as a gray dashed line. The uncertainties arising from the quark-mass uncertainty as well as from the quark-potential uncertainty are shown as a dark-blue and light-blue line respectively.}
\label{fig:mag_field}
\end{figure}

For each considered flavor (up, down and strange), we can compute the Zeeman-induced shift of the potential energy as
\begin{equation}\label{eq:deltau}
    \Delta U = U_{0}^- - U_{0}^+
\end{equation}
 with the use of Eq.~(\ref{eq:parameterscale}). Then---if the potential energy shift is electromagnetically dominated---we can combine equations (\ref{eq:electronzeem_true}), (\ref{eq:zeeman}) and (\ref{eq:bohrmagnetonquark}) to calculate the magnitude of the projection of the magnetic field $\vec{B}$ onto the direction of the flight of the nucleon (denoted as $B$). Results are shown in Fig.~\ref{fig:mag_field} along with errorbars arising from quark-mass uncertainty as well as from the quark-potential uncertainty.
 
The resulting magnetic fields are of the order $\sim 10^{13}$\,T. These would then be some of the strongest magnetic fields ever observed in nature (even surpassing those of record-breaking neutron stars dubbed magnetars~\cite{Kaspi:2017fwg}). 

Since magnetic fields inside nucleons are generated by quarks carrying orbital angular momentum (as well as by other phenomena such as by the addition of individual parton magnetic moments, etc.), to understand its origin and magnitude one would need to solve the proton-spin puzzle. This would mean to determine exactly how much orbital momentum is carried by which quark flavor at which energy. Even though this puzzle is not yet conclusively solved, we can try to get a rough estimate on the expected magnitude of the generated magnetic fields by using an analogy of a charge running in a loop defined by the dimensions of a proton. If we assume that a quark with a non-zero orbital angular momentum travelling at the speed of light, the induced magnetic field by its movement would be $B \sim 4 \cdot 10^{12}$\,T.

Of course, in reality, quark structure of a nucleon is much more complicated than this simplistic picture but it is used here as a demonstration of the possible expected magnetic fields to be found in nucleons.

\section{\label{sec:Summary} Discussion}
Zeeman-effect induced perturbation of the quark potential allowed us to compute the value of the projection of the magnetic field present at deep inelastic scattering events as B $\sim 8 \cdot 10^{12}$\,T in case that the helicity-dependent potential shifts are electromagnetically dominated.

Since quarks carry color-charge as well as electromagnetic one, they have an associated chromo-magnetic moment which would perturb the potential energy of a quark when exposed to a non-zero chromo-magnetic field. However, chromo-magnetic fields inside of a colorless object are on average zero due to the fact that they are not gauge invariant and one can have only non-zero variance of these fields. That is why color-magnetism would not contribute to the average potential shift that we observe in the statistical parton distributions. That being said, there are complex spin-spin and spin-orbit interactions that are mediated by the strong force that could be partially responsible for the potential splitting~\cite{Kovchegov_2016,Kovchegov_2019}. 

Since we are working with high-energy valence quark distributions, the value of the strong coupling constant would be large and one would not be able to use perturbative expansions of QCD to calculate the amount of their contribution. The estimate of the contribution of these effects to the potential splitting of quarks remains an open question that should be addressed in the future and this work presents an estimate of the classical magnetic field magnitude if that is the dominant driving force.

Even though these magnetic fields are strong, the energy stored in them in the volume of the nucleon cannot be larger than 1\,MeV (depending on the considered relative permeability of the medium), which is $\sim$ 0.1\% of the energy of the nucleon at rest. Fields of even bigger intensities have been found in heavy-ion collisions at STAR~\cite{Adam:2019mby}, which are generated by the Lorenz contraction of the electro-magnetic fields of the passing particles, whereas magnetic fields computed in this work are embedded in the parton distribution functions of the proton and describe its internal dynamics and properties.

Precise description of the magnetic properties of protons would be very useful for the understanding of the recently observed vortical structure of the "perfect fluid" of QCD that was observed in the global polarisation of the $\Lambda$ hyperons~\cite{STAR:2017ckg}. Their impact on the initial angular momentum distribution would furthermore significantly affect the outgoing polarization of hadrons in heavy-ion collisions~\cite{Becattini:2007sr}.

It would indeed be interesting to compute the helicity-dependent fits in the dimensionful framework directly to improve the accuracy of this work as well as to look for potential signatures of the quark-Zeeman effect in the rest frame of the proton (as for example in~\cite{Zavada:2007ww}) for learning more about the influence of the magnetic field on internal nucleon properties. It would also be of interest to estimate the contribution of the non-perturbative strong force contribution to the value of the quark-potential splitting.


\section{Acknowledgments}
I would like to thank Claude Bourrely, Petr Z\'{a}vada, Yuri Kovchegov, Peter Filip and Cyrille Marquet for discussion. I would also like to thank Jan Petr\r{u}, David Dob\'{a}\v{s} and David Wierichs for advice and support during this work.

This work has been supported from grant LTC17038 of the INTER-EXCELLENCE program at the Ministry of Education, Youth and Sports of the Czech Republic and the COST Action CA15213 THOR.
\bibliography{Biblio}

\end{document}